\documentclass[11pt,openany,openbib]{article}
\usepackage{amssymb}
\usepackage{amsmath}
\usepackage{epic}
\usepackage{eepic}
\usepackage{graphicx}
\DeclareSymbolFontAlphabet{\mathbb}{AMSb}
\usepackage{Vmargin}
\setmargins{1in}{1in}{6.5in}{9in}{0pt}{0pt}{12pt}{42pt}
\usepackage{times}
\usepackage{amsfonts}
\usepackage{eucal}
 
\newcommand{\qed}{\mbox{\rule{1.6mm}{4.3mm}}}

\newtheorem{theo}{Theorem}
\newtheorem{lm}[theo]{Lemma}
\newtheorem{co}[theo]{Corollary}

\newtheorem{defn}[theo]{Definition}

\begin{document}

\title{Exploration via design and the cost of uncertainty in keyword auctions} 
\author{ Sudhir Kumar Singh\\
UCLA \\
suds@ee.ucla.edu 
\and 
Vwani P. Roychowdhury\\
UCLA \& NetSeer Inc.\\
vwani@ee.ucla.edu
\and
 Milan Bradonji\'c \\
UCLA \\
milan@ee.ucla.edu
\and
 Behnam A. Rezaei\\
NetSeer Inc.\\
behnam@netseer.com
}
\maketitle

\begin{abstract}
We present a deterministic exploration mechanism for sponsored search auctions, which
enables the auctioneer to learn the {\it relevance} scores (Click-Through-Rates) of
advertisers, {\em and} allows  advertisers to estimate the true value of clicks generated
at the auction site. This exploratory mechanism deviates only minimally from the mechanism
being currently used by Google and Yahoo! in the sense that it retains the same pricing
rule, similar ranking scheme, as well as, similar mathematical structure of payoffs. In
particular, the estimations of the relevance scores and true-values are achieved by
providing a chance to lower ranked advertisers to obtain better slots. This allows the
search engine (the auctioneer) to potentially test a new pool of advertisers, and
correspondingly, enables new advertisers to estimate the value of clicks/leads generated
via the auction. Both these quantities are unknown {\em a priori}, and their knowledge is
necessary for the auction  to operate efficiently. We show that such an exploration policy
can be incorporated without any significant loss in revenue for the auctioneer.
We compare the revenue of the new mechanism to that of the standard mechanism (i.e.,
without exploration) at their corresponding {\it symmetric Nash equilibria}(SNE) and
compute the {\it cost of uncertainty}, which is defined as the relative loss in expected
revenue per impression. We also bound the loss in efficiency (i.e. social welfare), as
well as, in user experience due to exploration, under the same solution concept (i.e.
SNE). Thus the proposed exploration mechanism learns the relevance scores while
incorporating the incentive constraints from the advertisers who are selfish and are
trying to maximize their own profits, and therefore, the exploration is essentially
achieved via mechanism design. We also
discuss variations of the new mechanism such as truthful implementations.
\end{abstract}
\section{Introduction}
\label{intro}

\subsection{Preliminary Background}

With the growing popularity of web search for obtaining information, sponsored search
advertising, where advertisers pay to appear alongside the algorithmic/organic search
results, has become a significant business model today and is largely responsible for the
success of Internet Search giants such as Google and Yahoo!. In this form of advertising,
the Search Engine allocates the advertising space using an auction. Advertisers bid upon
specific keywords. When a user searches for a keyword, the search engine (the auctioneer)
allocates the advertising space to the bidding merchants based on their bid values and
quality scores/factors, and their ads are listed accordingly. Usually, the sponsored
search results appear in a separate section of the page designated as ``sponsored links''
above/below or to the right of the organic/algorithmic results and have similar display
format as the algorithmic results. Each position in such a list of sponsored links is
called a {\it slot}. Whenever a user clicks on an ad, the corresponding advertiser pays an
amount specified by the auctioneer. Generally, users are more likely to click on a higher
ranked slot, therefore advertisers prefer to be in higher ranked slots and compete for
them.

From the above description, we can note that after merchants have bid for a specific
keyword, when that keyword is queried, the auctioneer follows two steps. First, she
allocates the slots to the advertisers depending on their bid values. Normally, this
allocation is done using some {\it ranking function}. Secondly, she decides, through some
{\it pricing scheme}, how much a merchant should be charged if the user clicks on her ad
and in general this depends on which slot she got, on her bid and that of others. In the
auction formats for sponsored search, there are two ranking functions namely {\it rank by
bid} (RBB) and {\it rank by revenue}(RBR) and there are two pricing schemes namely {\it
generalized first pricing}(GFP) and  {\it generalized second pricing}(GSP) which have been
used widely. In RBB, bidders are ranked according to their bid values. The advertiser with
the highest bid gets the first slot, that with the second highest bid get the second slot
and so on. In RBR, the bidders are ranked according to the product of their bid value and
{\it quality score}. The quality score represents the merchant's relevance to the specific
keyword, which can basically be interpreted as the possibility that her ad will be viewed
if given a slot irrespective of what slot position she is given. In GFP, the bidders are
essentially charged the amount they bid and in GSP they are charged an amount which is
enough to ensure their current slot position. For example, under RBB allocation, GSP
charges a bidder an amount  equal to the bid value of the bidder just below her.

Formal analysis of such sponsored search advertising model has been done extensively in
recent years, from algorithmic as well as from game theoretic
perspective\cite{EOS05,MSVV05,Lah06,AGM06,Var06,LP07,MNS07}. In a formal setup, there are
$K$ slots to be allocated among  $N$ ($\geq K$) bidders. A bidder $i$ has a true valuation
$v_i$ (known only to the bidder $i$) for the specific keyword and she bids $b_i$. The
expected {\it click through rate} of an ad put by bidder $i$ when allocated slot $j$ has
the form $c_{i,j} = \gamma_j e_i$, i.e., separable into a position effect and an advertiser effect.
$\gamma_j$'s can be interpreted as the probability that an ad will be noticed when put in
slot $j$ and it is assumed that $\gamma_1 > \gamma_2 >\dots > \gamma_K > 0$. $e_i$ can be
interpreted as the probability that an ad put by bidder $i$ will be clicked on if noticed
and is referred to as the {\it relevance} of bidder $i$. This is the quality score used in
the RBR allocation rule mentioned earlier. The payoff/utility of bidder $i$ when given
slot $j$ at a price of $p$ is given by $e_i\gamma_j (v_i - p)$ and they are assumed to be
rational agents trying to maximize their payoffs.
Further, in typical slot auctions, bidders can adjust their bids up or down at any time and therefore
the auction can be viewed as a continuous-time process in which bidders learn each other's bids.
If the process stabilizes, the result can then be modeled as solution of the static one-shot game
of complete information, since each bidder will be playing a best-response to others' bids.

As of now, Google as well as Yahoo! use schemes that can be accurately  modeled as RBR
with GSP. The bidders are ranked according to $e_ib_i$ and the slots are allocated as per
these ranks. For simplicity of notation, assume that the $i$th bidder is the one allocated
slot $i$ according to this ranking rule, then $i$ is charged an amount equal to
$\frac{e_{i+1} b_{i+1}}{e_i}$. The revenue and incentive properties of this model has been
thoroughly analyzed in the above mentioned articles.

\subsection{The need for exploration}
In the standard model described above, it is implicitly assumed that the auctioneer knows
the relevance $e_i$'s, but in practice, this is not entirely true as new advertisers do
also join the game and the estimates for the advertisers getting lower ranked slots is
also generally poor as they hardly get any clicks. Further, it is also assumed that the
bidders know their true valuations accurately and bid accordingly, and high budget
advertisers and low budget  advertisers (e.g., mom-and-pop businesses) have similar
awareness and risk levels. In reality, an advertiser might not know her true value and
what to bid, and in particular a low budget  advertiser might be {\it
loss-averse}\cite{TK91} and may not be able to bid high enough to explore, due to the
potential risks involved. Furthermore, in the sponsored search auctions, the value is
 derived from the clicks themselves (i.e. rate of conversion or purchase given a
click), and therefore, unless she actually obtains a slot and receives user clicks, there
is essentially no means for her to estimate her true value for the keyword. Certainly, a
model that automatically allows one to estimate these key parameters (i.e. CTRs and true
values) is desirable.

\subsection{Results in this paper and related work}
Our goal in this paper is to study the problem of learning relevance scores and valuations
in a mechanism design framework while deviating only minimally from the mechanism
being currently used by Google and Yahoo!.
The problem of learning CTRs has also been
addressed in \cite{PO06,GP07,RDR07,IJMT05}. Our result is different from \cite{PO06} in
that the latter disregards the advertisers' incentives. The result in \cite{GP07} does
consider the advertisers' incentive; however, its goal is not to study exploration in the
mechanisms currently being used by search engines, but to implement a truthful mechanism
that also learns the CTRs, and therefore, it had to deviate from the current pricing
scheme. Our mechanism can also be easily adapted for truthful implementation via a new
pricing scheme, and in fact, all the revenue analysis remains the same as we shall discuss
later in the paper. Study in \cite{RDR07} is empirical and that in \cite{IJMT05} is not
exploration based, and restricts itself to a single slot case and does not consider
advertisers' incentives.

We recently learned about an independent study by Wortman et al. \cite{WVLL07} along lines
similar to ours, i.e., designing mechanisms for exploration that deviate minimally from
the standard model without exploration and then comparing their respective incentive
properties. Our mechanisms for exploration are, however, quite different and they
originated from a different set of approaches. Indeed, a preliminary draft that includes
all the main results presented in the current paper (although motivated a little
differently) was posted in early July 2007\cite{SBRR07}, well before the work in
\cite{WVLL07} was made publicly available. As discussed in greater detail in the
following, here are some of the distinctive features of our independent work: (i) Our
exploration mechanism is a deterministic one, unlike a randomized one analyzed in
\cite{WVLL07}; (ii) We explicitly discuss how advertisers could estimate their true
valuations under our exploration based mechanism. As argued before, true valuation is
often unknown a priori, and has to be accurately estimated; (iii) Besides studying the
loss in revenue due to exploration, we also explicitly discuss the loss in efficiency, as
well as, loss in user experience due to exploration; (iv) The tools and approaches used in
the analysis of our mechanism are very different from those presented in \cite{WVLL07},
and they highlight several interesting features of mechanism design and incentive
analysis. For example, we show that the {\em mathematical structure} of payoffs in our
exploration mechanism is the same as in the standard mechanism without exploration, which
allows us to utilize results from the latter. Thus, our approach represents an instance
where {\em reduction} among mechanism design problems is being successfully used as an
analytical tool.

Moreover, as we discuss later in Section~\ref{concl}, the problem of designing a family of
{\em optimal} exploratory mechanisms, which for example would provide the most information
while minimizing expected loss in revenue is far from being solved. The work in
\cite{WVLL07} and in this paper provide just two instances of mechanism design which do
provably well, but more work that analyze different aspects of exploratory mechanisms are
necessary in this emerging field. Thus, to the best of our knowledge, we are one of the
first groups to formally study the problem of estimating relevance and valuations from
incentive as well as learning theory perspective without deviating much from the current
settings of the mechanism currently in place.

In the following we summarize our results as well as the organization of the rest of the
paper:

\begin{enumerate}
\item We design a deterministic exploration mechanism to learn the relevance scores by deviating
minimally from the mechanism being currently used by Google and Yahoo! in the sense that
it retains the same pricing rule, as well as, similar ranking scheme. In particular, the
estimation of the relevance scores is achieved by providing a chance to lower ranked
advertisers to obtain better slots. Qualitatively, some top slots are designated for
exploration purposes and each of the advertisers whose relevance is to be estimated, is
given an equal chance to appear in those slots. In Section \ref{Exp-GSP}, we formally
introduce this exploration mechanism which we call {\bf Exp-GSP} and the standard RBR with
GSP mechanism without exploration is referred to as {\bf GSP}.

\item In Section \ref{incentive},
we study the incentive properties of {\bf Exp-GSP} mechanism by modeling it as one shot
static game of {\it complete information}, like in the case of {\bf
GSP}\cite{EOS05,Var06}. We show that the mathematical structure of the payoffs of the
bidders in {\bf Exp-GSP} is the same as in {\bf GSP}, and therefore all the incentive
analysis from {\bf GSP} can be adopted for {\bf Exp-GSP}. This further corroborates our
claim that our exploration mechanism deviates only minimally from {\bf GSP} and indeed our
approach can also be understood as reduction among mechanism design problems. {\em
Furthermore,} another interesting feature of our exploration mechanism is that the
attention or the quality of service (in terms of position based CTRs i.e. probability of
being noticed) provided to advertisers is still in the same relative order as in standard
mechanism without exploration.

\item It is clear that any exploration mechanism will incur some cost in terms of revenue
compared to the case when we do not need an exploration. We formalize this cost via {\it
cost of uncertainty} which  is defined as the relative loss in expected revenue of the
auctioneer per impression. To this end, we compare the revenue of the  {\bf Exp-GSP} to
that of {\bf GSP} at their corresponding {\it symmetric Nash equilibria}(SNE) and bound
the {\it cost of uncertainty}. Our analysis confirms the intuition that a higher cost is
incurred for better exploration i.e. there is a tradeoff between quality of
exploration/estimation and the revenue. Nevertheless, the associated parameters can be
tuned to ensure a suitable balance between these two conflicting needs- minimizing the
loss in revenue while allowing for sufficient exploration to be able to estimate
parameters such as the relevance scores. These revenue properties are studied in the
Section \ref{revsec}.

\item Section \ref{effsec} discusses the loss in efficiency in {\bf Exp-GSP}
compared to {\bf GSP}. As in the case of revenue,
there is a tradeoff between efficiency (i.e. social welfare) and the quality of exploration/estimation.
Additionally, our analysis also suggests that
closer we are to the optimal efficiency
(i.e. the case when the auctioneer knows true values of relevance scores and the advertisers know that of their
valuations), lesser we lose in the efficiency due to exploration. This means that
during several phases of the exploration the loss in the efficiency degrades.
Similar observations can also be obtained for {\it user experience} which can be
defined as the total clickability of all ads.

\item  In Section \ref{est}, we discuss how our exploration mechanism i.e. {\bf Exp-GSP} can be
used to estimate relevance scores and
valuations, as well as, the quality of such estimation using Chernoff bound arguments.

\item In all the Sections from \ref{Exp-GSP} through \ref{est}, we
restrict ourselves to a standard assumption in literature that the CTRs are separable. In
Section \ref{var}, \emph{we remove this assumption} and study some other variations of
{\bf Exp-GSP}. In particular, by imposing a new pricing rule we can turn our exploration
mechanism to a truthful one. Moreover, a similar upper bound on the cost of uncertainty is
established as in the case of {\bf Exp-GSP} with separable CTRs.
\end{enumerate}

\section{An exploration based Generalized Second Price mechanism }
\label{Exp-GSP}
In this section, we formally introduce our exploration mechanism.
First we setup some notations and definitions.

{\bf Notation:}
There are $N$ advertisers/bidders bidding for a specific keyword and this keyword appears several times during a day.
There are $K \leq N$ slots to be allocated among the bidders for this keyword. A bidder $i$ has a true valuation
$v_i$ for this keyword and she bids $b_i$. The
expected {\it click through rate} of an ad put by bidder $i$ when allocated slot $j$ has
the form $CTR_{i,j} = \gamma_j e_i$, i.e., separable into a position effect and an advertiser effect
wherein
$e_i$ is the {\it relevance} of the bidder $i$. Further, it is assumed
that $\gamma_j > \gamma_{j+1}$ for all $j=1,2,\dots,K$ and $\gamma_j =0$ for all $j > K$.
The search engines' estimate of relevance $e_i$ of bidder $i$ is denoted by $q_i$ and
bidder $i$'s estimate of her relevance $e_i$ is denoted by $f_i$.
There are no budget constraints.

{\bf Explore slots and tuning parameters:}
Auctioneer chooses two parameters $n \leq N$ and $L \leq K$.
Auctioneer designates top $L$ slots for exploratory purpose. Let us call these slots as {\it explore slots}
and slots $L+1$ through $K$ will be called {\it non-explore}.
Auctioneer decides a set $F$ of $n$ bidders whose relevance, she wants to estimate.
As described in the mechanism below, these $n$ bidders will be the top $n$ bidders according to
auctioneer's ranking rule. If auctioneer wants to just improve the estimate for some bidders, she chooses $n \leq K$ and
if she also wants to estimate the relevance of some new bidder or some left-out bidder, she chooses $n \geq K+1$.
The parameters $n$ and $L$ are publicly known. Further, as we shall see below,
the mechanism has $n$ steps and during these $n$ steps,
the bidders in set $F$ will be given equal chance
to appear in the {\it explore slots} in the sense that they appear exactly once in each explore slot.
During a step, when a bidder does not appear in one of the explore slots, she competes for {\it non-explore} slots
with all the bidders who  do not appear in the explore slots. Now we are ready to formally describe the new
mechanism which we call {\bf Exp-GSP} ({\bf Exp}loratory-{\bf G}eneralized {\bf S}econd {\bf P}rice).  \\

{\bf The Exp-GSP Mechanism:}

\begin{itemize}

\item Bidders report their bids $b_1, b_2, \dots, b_N$.

\item {\bf Ranking Bidders:} Auctioneer uses {\bf RBR} to rank the bidders i.e. she ranks the bidders in the decreasing order of $q_i b_i$.
For clarity of notation, let us rename the bidders according to this ranking, i.e., bidder $m$ is the one ranked $m$
in this ranking.

\item {\bf Allocating Explore Slots:} There are $n$ steps in the mechanism and the $n$ bidders in $F$ are ordered in each step as follows.
The ordering at step $1$ is the above mentioned RBR ranking i. e.  $[1, 2, \cdots, L \mid (L+1), \cdots,n]$.
This order is cyclicly shifted towards left for $n-1$ more steps.
Thus the ordering in step $2$ is $[2, 3, \cdots, L, (L+1) \mid (L+2), \cdots, n, 1]$ and that in step $3$ is
$[3, \cdots (L+2) \mid (L+3),\cdots, n, 1, 2]$ and so on.
In a particular step, for $ j \leq L$, the $j$th slot is assigned to the bidder having rank $j$ in this
cyclicly rotating ordering at that step.  For example, in step 1, the slot $j\leq L$ is allocated to the bidder $j$.
In step $2$, the slot $j\leq L$ is allocated to the bidder $j+1$ and in step $n$, first slot is assigned to the bidder $n$,
and for $2 \leq j \leq L$, the $j$th slot is allocated to the bidder $j-1$.
In a particular step, a bidder will be called
{\it explore-active} if she is assigned one of {\it explore} slots in that step.
Note that this cyclicly shifting rule ensures that during total of $n$ steps, each of the $n$
bidders in $F$ gets to each explore slot {\it exactly once}, thus each one is explore-active for exactly
$L$ steps out of the $n$ steps. Also, in each step there are exactly $L$ {\it explore-active} bidders.

\item {\bf Allocating non-Explore Slots:} Bidders from $F$ who are {\em not} explore-active at a particular
step along with bidders not in $F$, are allocated to non-explore slots as follows.
Let $i_1 < i_2 < \dots < i_{N-L}$ be the bidders who are not explore-active in this particular step.
Recall that we renamed the bidders according to the RBR ranking. Now the slot $L+j$ for $1 \leq j \leq K-L$ is assigned to the bidder
$i_j$. For example, in step $1$, we have $i_j = L+j$; in step $2$ we have $i_1 =1$ and $i_j = L+j$ otherwise, and in step $n$ we have
$i_j = L+j-1$.

\item {\bf Payments based on GSP :} A bidder $i$ is charged an amount equal to $\frac{q_{i+1}b_{i+1}}{q_i}$ {\it per-click}.

\end{itemize}

{\bf Nomenclatures:}
For the rest of the paper, we fix some nomenclatures. The standard one step mechanism with RBR ranking and GSP pricing
will be referred to as {\bf GSP} and the new exploration based mechanism described above ( all the $n$ steps together) will be
referred to as {\bf Exp-GSP}.
Further, we will refer $\gamma_j$'s to as {\it position based cilck-through rates}.
Let  $I_i$ denote  all the information about the bidder $i$ i.e. $I_i$ includes
bidder $i$'s true relevance $e_i$, auctioneer's estimate of her relevance $q_i$, her estimate of her relevance $f_i$,
her true value $v_i$ and her estimate of her true value $\tilde{v}_i$, all the knowledge of bidder $i$ about the auction game etc.
An instance of the {\bf GSP} is represented by $(N, K, (\gamma_j), (I_i))$ and that of {\bf Exp-GSP} by
$(N, K, n, L, (\gamma_j), (I_i))$.
Clearly, any given instance  $(N, K, (\gamma_j), (I_i))$ of {\bf GSP} is equivalent to an instance
$(N, K, n, L, (\gamma_j), (I_i))$ of {\bf Exp-GSP} where $n=1,L=0$.
Further, as we show in Section \ref{incentive}, a large class of instances of {\bf Exp-GSP} of our interest can also be mapped to instances of
{\bf GSP}
with properly defined position based click-through rates.
This corroborates our claim that we deviate
minimally from the mechanism currently in place.

\section{Incentive properties}
\label{incentive}

In this section, we study the incentives properties of $n$-step {\bf Exp-GSP} mechanism modeling it as one shot static game of
{\it complete information}, where the advertisers know others' bids, and play the best
response to others' bids given their {\it current} estimates of their CTR's and their true valuations.
This is reasonable as the bidding process can be thought of as a
continuous process, where bidders learn each other's bids\cite{EOS05,Var06,Lah06,LP07}.
As we explain in the following, a large class of the instances of {\bf Exp-GSP} can be mapped to instances of {\bf GSP}
with properly defined click-through rates and therefore will allow us to use the results on {\bf GSP}.
This corroborates our claim that we deviate
minimally from the mechanism currently in place.
The solution concept we will use is {\em Symmetric Nash Equilibria(SNE)/locally envy-free equilibria} studied in \cite{EOS05,Var06}.
First, we define {\it effective CTR } which will help us mapping instances of {\bf Exp-GSP} to that of {\bf GSP}.

\begin{defn}
{\bf Effective Click-Through Rates:}
Let $l_1, l_2, \dots, l_n$ be the slot positions that a bidder $j$ is assigned in the steps $1,2,\dots, n$ of {\bf Exp-GSP} respectively, then
the effective CTR of a bidder $i$ for slot $j \leq N$ denoted as $\tilde{c}_{i,j}$ is defined as $\sum_{m=1}^n c_{i,l_m}$.
Thus for the separable case, the {\bf effective position based CTR} for slot $j \leq N$ denoted $\theta_j$ is $\sum_{m=1}^n \gamma_{l_m}$.
\end{defn}

Intuitively, the effective CTR of a bidder $i$ for slot $j$
is the sum of the expected CTR of bidder $i$ for each of the $n$ step in {\bf Exp-GSP} if he would have been ranked $j$.
It is not hard to derive the following lemma.

\begin{lm}
\label{epctr}
Let $\gamma = \sum_{j=1}^L \gamma_j$  then
 \begin{equation}
\label{thetas0}
\theta_m=\left\{\begin{array}{ll}
\gamma+d_m & \textrm{ if $m \leq n$} \\
n \gamma_m & \textrm{ if $m > n$} \\
\end{array} \right.
\end{equation}
where
\begin{align}
\label{dms}
d_{m} = \left\{\begin{array}{l}
(n-L-(m-1))\gamma_{L+m} +  \\
\gamma_{L+1}+\gamma_{L+2}+\cdots+\gamma_{L+m-1} \textrm{ {\bf if $m \leq L$}} \\
\\
(m-L)\gamma_m +\gamma_{m+1} + \cdots + \gamma_{m+L-1}+ \\
 (n-m-L+1) \gamma_{m+L}  \textrm{ {\bf if $L \leq m \leq n-L$}} \\
\\
(m-L) \gamma_m + \\
\gamma_{m+1}+\cdots+\gamma_n  \textrm{  {\bf if $m \geq n-L $}} \\
\end{array}
\right.
\end{align}
\end{lm}

In the above lemma, $\gamma$ basically represents the effective position based click through that a
bidder obtains from the explore slots (in $n$ steps) and $d_m$ represents the
effective position based click through that the bidder $m$ obtains from the non-explore slots (in $n$
steps). In particular, the $d_m$ indicates how many steps the bidder $m$ spends in
specific non-explore slots. For example, $d_1=(n-L)\gamma_{L+1}$ indicates that the bidder
$1$ spends $(n-L)$ steps in the slot numbered $(L+1)$, $d_2=(n-L-1)\gamma_{L+2}+
\gamma_{L+1}$ indicates that the bidder $2$ spends $(n-L-1)$ steps in the slot $(L+2)$ and
one step in the slot $(L+1)$, and so on for other bidders. In the following lemma we
observe that these effective position based CTRs are in fact strictly monotonically decreasing like
$\gamma_j$'s. The proof is provided in the Appendix.

\begin{lm}
\label{thetas}
Let $\tilde{K} = \max\{K, n\}$, $ n \leq \min\{K + 1,K+L\}$, and $L \leq \frac{1}{2} (n-1)$ then
\begin{displaymath}
\theta_1 > \theta_2 \dots > \theta_{\tilde{K}} > 0
\end{displaymath}
and $\theta_{i} = 0 $ for all $i > \tilde{K}$.
\end{lm}

Now under {\bf Exp-GSP} the payoff of the bidder $m$ is
\begin{equation}
u_m =\theta_m e_m (v_m -\frac{q_{m+1} b_{m+1}}{q_i}).
\end{equation}
which has exactly the same functional form as in {\bf GSP} where
$\theta_m$'s takes the place for $\gamma_m$'s and therefore our name for $\theta_m$'s makes sense.
Thus an instance $(N, K, n, L, (\gamma_j), (I_i))$ of {\bf Exp-GSP}  where $ n \leq K + 1$, and $L \leq \frac{1}{2} (n-1)$,
can be mapped to an
instance $(N, \max\{K,n\}, (\theta_j), (I_i))$ of {\bf GSP}.
We formalize this in the following theorem.

\begin{theo}
\label{exgsp-to-gsp}
For each instance $(N, K, n, L, (\gamma_j), (I_i))$ of {\bf Exp-GSP} with $ n \leq K + 1$, and $L \leq \frac{1}{2} (n-1)$,
there is an instance  $(\tilde{N}, \tilde{K}, (\tilde{\gamma}_j), (\tilde{I}_i))$ of {\bf GSP}
such that the game induced by $(N, K, n, L, (\gamma_j), (I_i))$ is equivalent to
the game induced by $(\tilde{N}, \tilde{K}, (\tilde{\gamma}_j), (\tilde{I}_i))$.
In particular, $ \tilde{N}=N, \tilde{K} =\max\{n,K\}, \tilde{\gamma}_j=\theta_j, \tilde{I}_i=I_i$ where $\theta_j$'s are defined
by Equations \ref{thetas0} and \ref{dms}.
\end{theo}

It is interesting to note that even though we allowed lower ranked bidders to obtain top slots,
the competition for the {\bf non-explore slots keeps the effective position based CTRs still in the same relative
order.} The highest ranked bidder still gets the best service compared to others although
her effective payoff might have decreased. A lower ranked bidder still gets relatively
lower quality of service than the bidders above her although her payoff might have
improved.
This same structural form of payoffs allows us to derive Theorem \ref{exgsp-to-gsp} and therefore
to utilize the results on {\bf GSP}
studied in \cite{EOS05,Var06,Lah06, LP07, BCPP06, AGM06} and in particular the
following theorem on existence of pure Nash equilibria for {\bf Exp-GSP}. Thus our approach can also be
understood as reduction among mechanism design problems.

\begin{theo}
There always exist a pure Nash equilibrium bid profile for the  {\bf Exp-GSP}.
\label{expgspnash}
\end{theo}

As noted in the above theorem, there always exist pure strategy Nash equilibria
for the {\bf Exp-GSP} auction game. However, this
existential proof does not give much insight about what equilibria might arise in practice.
Edelmen et al \cite{EOS05} proposed a class of Nash equilibria which they call as {\it
locally envy-free equilibria} and argue that such an equilibrium arises if agents are
raising their bids to increase the payments of those above them, a practice which is
believed to be common in actual keyword auctions. Varian\cite{Var06} independently
proposed this solution concept which he calls as {\it symmetric Nash equilibria(SNE)} and
provided some empirical evidence that the Google bid data agrees well with the SNE bid
profile. In a similar way we can obtain the following observation.

\begin{theo}
\label{snebp}
An {\bf SNE} bid profile $b_i$'s for {\bf Exp-GSP} satisfies
\begin{align}
(\theta_i - \theta_{i+1}) v_{i+1} q_{i+1} + \theta_{i+1} q_{i+2} b_{i+2} \leq \theta_{i} q_{i+1} b_{i+1} \nonumber \\
\leq (\theta_i - \theta_{i+1}) v_{i} q_{i} + \theta_{i+1} q_{i+2} b_{i+2}
\end{align}
for all $ i=1,2,\dots, N$.
\end{theo}

Note that the Theorem \ref{snebp} assumes that the bidders know their true valuations $v_i$'s, however
the theorem holds evenif it is not the case by replacing $v_i$ by bidder $i$'s current estimate of
her true valuation.

Now, recall that in the {\bf Exp-GSP}, the bidder $i$ pays an amount $\frac{ q_{i+1} b_{i+1} }{q_i}$ per-click,
therefore the expected payment $i$ makes under {\bf Exp-GSP} (in $n$ steps) is
$\theta_i e_i \frac{ q_{i+1} b_{i+1} }{q_i} = \frac{e_i}{q_i} \theta_i q_{i+1} b_{i+1}$.
Thus the best {\bf SNE} bid profile for advertisers (worst for the auctioneer) is minimum bid profile
possible according to Theorem \ref{snebp} and is given by
\begin{align}
\label{minsne}
 \theta_{i} q_{i+1} b_{i+1} = \sum_{j=i}^{\tilde{K}} (\theta_j - \theta_{j+1}) v_{j+1} q_{j+1}.
\end{align}

For the revenue comparison in the next section, we fix this minimum SNE bid profile as the solution concept.
The same result essentially hold for the maximum SNE  bid profile as well.

\section{Revenue comparison and the cost of uncertainty}
\label{revsec}
In this section we study the
revenue properties of {\bf Exp-GSP} and compare it to that of {\bf GSP}.
We first define the {\it cost of uncertainty} to formalize the loss of revenue due to exploration.

\begin{defn}
{\bf Cost of uncertainty:} Let $R_0$ be the expected revenue of the auctioneer for {\bf GSP} at its minimum SNE and
$R$ be her expected revenue for {\bf Exp-GSP} at the corresponding minimum SNE, then ``cost of uncertainty''
associated with the exploration is defined as {\bf $\frac{R_0 - \frac{1}{n} R}{R_0}$} i.e. the expected
{\bf relative loss in the revenue per impression} and
is denoted as $\rho$.
\end{defn}

Using Equation \label{minsne}, we have
\begin{displaymath}
R_0 = \sum_{s=1}^K \sum_{j=s}^K \frac{e_s}{q_s}  (\gamma_j - \gamma_{j+1})q_{j+1} v_{j+1}
\end{displaymath}
and
\begin{displaymath}
R =  \sum_{s=1}^{\tilde{K}} \sum_{j=s}^{\tilde{K}} \frac{e_s}{q_s} (\theta_j - \theta_{j+1}) q_{j+1} v_{j+1}
\end{displaymath}

\begin{displaymath}
\therefore R_0- \frac{1}{n} R = \sum_{s=1}^{\tilde{K}} \sum_{j=s}^{\tilde{K}} \frac{e_s}{q_s}  \left[(\gamma_j - \gamma_{j+1}) - \frac{1}{n}
(\theta_j - \theta_{j+1})\right] q_{j+1} v_{j+1}.
\end{displaymath}

By utilizing the relationship among $\gamma_j$'s and $\theta_j$'s we can
obtain the following theorem which provides a nice upper bound on the {\it cost of uncertainty}.
The proof this theorem is provided in the Appendix.

\begin{theo}
\label{cou}
Let $R_0^l$ be the revenue of auctioneer from top $l$ bidders and $R_0$ be her total revenue in {\bf GSP} and let
\begin{equation}
\label{c}
c= \min_{1 \leq j < n-L} \frac{\gamma_{j+L} - \gamma_{j+1+L}}{\gamma_j -\gamma_{j+1}}.
\end{equation}
then
\begin{align}
\rho(L,n) \leq  \left\{ 1 - \min\{1,c\} (1-\frac{2L}{n}) \right\} \left(\frac{R_0^{\min\{n,K\}}}{R_0}\right) \nonumber\\
\leq \left\{ 1 - \min\{1,c\} (1-\frac{2L}{n}) \right\}.
\end{align}
\end{theo}

First, note that the above bound is $0$ when $L=0$, indicating no revenue loss when there is no exploration.
Further, given an $n$, as $L$ increases the bound deteriorates confirming our intuition that higher cost is incurred
for better exploration. Also for a given $L$, we can note that the factor $\frac{R_0^{\min\{n,K\}}}{R_0}$
is dominant and increases as $n$ increases and therefore the bound deteriorates as $n$ increases.
We see that auctioneer can tune parameters $L$ and $n$ so as to
improve revenue, smaller the $L$ and $n$, better off the auctioneer is. But as the auctioneer also
wants to get some valuable information so as to estimate parameters such as relevance of
the advertisers and do also want to give flexibility to lower ranked bidders to figure out
their valuations, she would like to keep $L$ and $n$ to be large. Therefore, the auctioneer
can choose a suitable $L$ and $n$ to balance between these two conflicting needs.
Furthermore, it is clear that a finer analysis will reveal much better revenue guarantee i.e. even smaller $\rho$.
For example, usually the expression on right hand side of Equation \ref{c} in the above theorem is dominated
by $j=1$, however if we look at the expression for revenue the $j=1$ term appears only once
unlike all other $j$'s and neglecting $j=1$ does not noticeably change the difference in the revenues
and therefore a better $c$ might be achievable with  this fine tuning. \\

We can also note that Theorem \ref{cou} still holds true when we replace the RBR ranking rule
in {\bf GSP} and {\bf Exp-GSP} by any weighted ranking rule (i.e. in the decreasing order of $w_ib_i$'s)
and change the payment rules accordingly (i.e. $\frac{w_{i+1}b_{i+1}}{w_i}$ per-click to the $i$th ranked bidder).

\section{Efficiency comparison}
\label{effsec}
Revenue is a natural yardstick for comparing different auction forms from the viewpoint of
the seller (the auctioneer), however from a social point of view yet another yardstick
that is natural and may be important is {\it efficiency}, that is, the social value of the
object. The object should end up in the hands of the people who value it the most. The
efficiency in the adword auction model is therefore the total valuation, and turns out to
be the  combined profit of the auctioneer and all the bidders. Let us denote the
efficiency for the {\bf Exp-GSP} as $E$ and that for  {\bf GSP} as $E_0$ then,
\begin{align}
E= \sum_{m=1}^{\tilde{K}} \theta_m e_m v_m \\
E_0 = \sum_{m=1}^K  \gamma_m e_m v_m .
\end{align}

Using Lemma \ref{epctr} and rearranging the terms in $E$ we get,

\begin{lm}
\label{effre}
\begin{align}
E = \sum_{m=1}^K \gamma_m y_m   \\
\end{align}
 where
\begin{align}
y_m = \left\{ \begin{array}{l}
\sum_{i=1}^n e_iv_i   \textrm{      if   } m \leq L \\
\\
(n-m+1) e_{m-L} v_{m-L} + \sum_{i=m-L+1}^{m-1} e_iv_i \\
 + (m-L) e_m v_m  \textrm{  if  } L < m \leq n \\
\\
n e_m v_m \textrm{  if  } m > n \\
\end{array} \right.
\end{align}
\end{lm}

The above lemma allows us to bound the loss in efficiency due to exploration as we note in the following theorem
whose proof is deferred to Appendix.

\begin{theo}
\label{effloss}
Let $E_0^{e} = \sum_{i=1}^L \gamma_m e_m v_m$, $E_0^{ne} = \sum_{i=L+1}^{n} \gamma_m e_m v_m$ then
the relative loss in efficiency per impression is
\begin{align}
\frac{E_0 - \frac{1}{n} E}{E_0} \leq  \left\{(1-\beta) \left(\frac{E_0^e}{E_0}\right) + \eta \left(\frac{E_0^{ne}}{E_0}\right)\right\}
\end{align}
where
\begin{align}
\beta = \frac{1}{n}\frac{\sum_{i=1}^n e_i v_i}{\max_{1 \leq m \leq L} e_mv_m} , \eta = \max_{L < m \leq n} \left\{\max_{m-L \leq i \leq m} \left(1-\frac{e_iv_i}{e_mv_m}\right)\right\}.
\end{align}
\end{theo}

First, note that the above bound is $0$ when $L=0$, indicating no efficiency loss when there is no exploration.
Further, given an $n$, as $L$ increases the bound deteriorates and similarly
for a given $L$, the bound deteriorates as $n$ increases. Apart from the tuning parameters
$n$ and $L$, note that there is another interesting parameter $\eta$ which actually depends on the true
relevance and the true values of the advertisers. In particular, it indicates that how far
the current estimates are from the true ones. For example, in the extreme case when the auctioneer knows the true relevances,
then the ordering by $q_mv_m$, will be equivalent to the ordering by $e_mv_m$ and $\eta$ will infact  be  $0$,
improving the bound. Thus closer we are to the optimal efficiency, lesser we lose in efficiency due to exploration.
 The proof of Theorem \ref{effloss} includes the following observation in the case when
the ordering by $q_mv_m$ is same as the ordering by $e_mv_m$.

\begin{co}
Under the assumption that $e_mv_m \geq e_{m+1}v_{m+1}$ for all $1 \leq m \leq n$ the upper bound in Theorem \ref{effloss} can be improved to
\begin{align}
  \left\{(1-\alpha) \left(\frac{E_0^e}{E_0}\right) - \frac{L}{n} \omega \left(\frac{E_0^{ne}}{E_0}\right)\right\} \nonumber\\
\nonumber \\
\textrm{where   }  \alpha = \frac{1}{n} \frac{\sum_{i=1}^n e_iv_i}{e_1v_1},
\omega = \min_{L < m \leq n} \left( \frac{e_{m-1}v_{m-1}}{e_m v_m} - 1 \right) . \nonumber
\end{align}
\end{co}

Now let us consider the effect on the {\it user experience} due to exploration.
Following \cite{LP07}, the {\it user experience} can be defined as the total clickability of all the ads i.e. how likely
an user is to click on the ads altogether.
Therefore, for {\bf GSP} it is $\sum_{m=1}^K \gamma_m e_m$ and that for {\bf Exp-GSP} it is $\sum_{m=1}^{\tilde{K}} \theta_m e_m$.
Clearly, similar observations in the loss of {\it user experience} due to exploration can be obtained as in the case
of efficiency.

\section{Estimating the relevance and valuations }
\label{est}

Let $M_i$ be the number of clicks that the advertiser $i$ receives in {\bf Exp-GSP} then
her relevance $e_i$ is estimated as $\frac{M_i}{\theta_i}$ and the deviation will
not be high as can be argued using Chernoff bound arguments.
Formally, let $M_{i,j}$ be a $0-1$ random variable indicating whether the advertiser $i$ gets
a click in the $j$th impression (i.e. $j$th step in {\bf Exp-GSP}) or not and $M_i=\sum_{j=1}^n M_{i,j}$. Clearly,
$E[M_i]= \sum_{j=1}^n E[M_{i,j}] = \theta_i e_i$.  Then by Chernoff bound, for any $0< \delta < 1$, we have
\begin{align}
Pr(|e_i - \frac{M_i}{\theta_i}| \geq \delta e_i) \leq 2 e^{-\theta_i e_i \frac{\delta^2}{3}} .
\end{align}

A simple calculation implies that, we can get an estimate of $e_i$ within a $\delta$ fraction with probability $1-\epsilon$
as long as we have,
\begin{align}
\theta_i \geq \frac{3}{\delta^2 e_i} ln(\frac{\epsilon}{2}).
\end{align}

Normally we will be interested in estimating the relevance of lower ranked advertisers and clearly for
them the value of $\theta_i$ increase as we increase the value of $L$ and we can guarantee a better estimation.
In particular, given a value of $L$ and $n$, we can have reliable estimation with probabilty  $1-\epsilon$
within a fraction of $\sqrt{\frac{3}{e_i\theta_i} ln(\frac{2}{\epsilon})}$ and an additive estimation within
 $\sqrt{\frac{3}{\theta_i} ln(\frac{2}{\epsilon})}$.
The above estimation can be improved even further by sampling from many phases of {\bf Exp-GSP}.
Note that even if we consider the $l$ phases of {\bf Exp-GSP} as a single shot game,
the results of the sections \ref{incentive} and \ref{revsec} remains unchanged
and in particular the {\it cost of uncertainty} does not change.
As above using Chernoff-bounds arguments, we can obtain an additive estimation within $\delta$ with probability $1-\epsilon$
if we use $l$ phases where
\begin{align}
l \geq \frac{3}{\delta^2 \theta_i} ln(\frac{\epsilon}{2}).
\end{align}
Thus we can obtain an estimation negligibly (i.e. inverse polynomially in parameter $n,L$) close to the true value
with probability exponentially close to $1$ in polynomially many phases of  {\bf Exp-GSP}. We summarize the above observation in
the following theorem.

\begin{theo}
The relevance of the advertiser $i$ can be estimated within $\delta$ with probability $1-\epsilon$
 by using $l$ phases of {\bf Exp-GSP} where,
\begin{align}
l \geq \frac{3}{\delta^2 \theta_i} ln(\frac{\epsilon}{2}). \nonumber
\end{align}
Even a single phase of {\bf Exp-GSP} can provide pretty good estimate
with probabilty  $1-\epsilon$
within $\sqrt{\frac{3}{\theta_i} ln(\frac{2}{\epsilon})}$ of her true relevance.
\end{theo}

In a similar way, the advertisers can estimate their valuations.
A reasonable way an advertiser can estimate her value is via tracking conversions i.e.
which clicks lead to a purchase or an activity of the advertiser's interest.
Let $x_i$ be the value advertiser $i$ derives from a single conversion and $a_i$ be the conversion probability per click
and $Q_i$ be the total number of conversions she obtains in {\bf Exp-GSP}  then she can
estimate her value to be $\frac{Q_i}{\theta_i\tilde{f}_i} x_i$ per click and using Chernoff-bound as above and
union bound we can
argue that this estimation is very good. Here $\tilde{f}_i$ is her updated estimate of her relevance using
the current phase of {\bf Exp-GSP}.
In reality, it might be difficult to track conversions but it is not clear how can the advertiser estimate
without the knowledge of her conversion rate.
Further, it is also possible that she derives some values from impressions and clicks even though it does not lead
to a conversion. For example, an impression gives some branding value and a click improves her relevance score
even when they do not lead to a conversion. In this general case,
let $x_i^I, x_i^C,x_i^A$ be the values advertiser $i$ derives from an impression, a click and a conversion respectively then
she can estimate her value to be $\frac{nx_i^I+ M_ix_i^C+Q_ix_i^A }{\theta_i\tilde{f}_i}$ per click.

\section{Variations of Exp-GSP: \newline {\small
Truthful Implementation and non-separable Click-through rates}}
\label{var}
Recall from Section \ref{incentive} that the effective CTR of a bidder $i$ for slot $j$ denoted $\tilde{c}_{i,j}$
is the sum of the expected CTR of bidder $i$ for each of the $n$ step in {\bf Exp-GSP} if he would have been ranked $j$
and in a similar way as for $\theta_i$'s we can derive the following lemmas.

\begin{lm}
\label{ectr}
Let $\beta_i = \sum_{j=1}^L c_{i,j}$  then
 \begin{equation}
\label{newctr}
\tilde{c}_{i,m} =\left\{\begin{array}{ll}
\beta_i +d_{i,m} & \textrm{ if $m \leq n$} \\
n c_{i,m} & \textrm{ if $m > n$} \\
\end{array} \right.
\end{equation}
where
\begin{align}
\label{dims}
d_{i,m} = \left\{\begin{array}{l}
(n-L-(m-1)) c_{i,L+m} +  \\
c_{i,L+1}+c_{i,L+2}+\cdots+c_{i,L+m-1} \textrm{ {\bf if $m \leq L$}} \\
\\
(m-L)c_{i,m} +c_{i,m+1} + \cdots + c_{i,m+L-1}+ \\
 (n-m-L+1) c_{i,m+L}  \textrm{ {\bf if $L \leq m \leq n-L$}} \\
\\
(m-L) c_{i,m} + \\
c_{i,m+1}+\cdots+c_{i,n}  \textrm{  {\bf if $m \geq n-L $}} \\
\end{array}
\right.
\end{align}
\end{lm}

\begin{lm}
\label{newcijs}
Let $\tilde{K} = \max\{K, n\}$, $ n \leq \min\{K + 1,K+L\}$, and $L \leq \frac{1}{2} (n-1)$ then for all $1 \leq i \leq N$
\begin{displaymath}
\tilde{c}_{i,1} > \tilde{c}_{i,2} \dots > \tilde{c}_{i,\tilde{K}} > 0
\end{displaymath}
and $\tilde{c}_{i,j} = 0 $ for all $j > \tilde{K}$.
\end{lm}

Consider any ranking based mechanism and the corresponding exploration based generalization as described
in Section \ref{Exp-GSP} with payment rule modified accordingly then the instances of the two mechanisms are
given by $(N, K, (c_{i,j}), (I_i))$ and $(N, K, n, L, (c_{i,j}), (I_i))$ respectively. Therefore, using the Lemmas \ref{ectr}, \ref{newcijs}
we can obtain a reduction similar to Theorem \ref{exgsp-to-gsp}:
for each instance $(N, K, n, L, (c_{i,j}), (I_i))$ of exploration based mechanism with $ n \leq K + 1$, and $L \leq \frac{1}{2} (n-1)$,
there is the instance  $(N, \max\{n,K\}, (\tilde{c}_{i,j}), (I_i))$ of corresponding one step mechanism without exploration
such that the game induced by $(N, K, n, L, (c_{i,j}), (I_i))$ is equivalent to
the game induced by $(N, \max\{n,K\}, (\tilde{c}_{i,j}), (I_i))$, where $\tilde{c}_{i,j}$ is given by the Equations \ref{newctr}, \ref{dims}.
Therefore, we can use all the results from one step mechanism without exploration.
In the following we consider two variations of {\bf Exp-GSP} -
(i) for the given ranking mechanism the goal is to design a truthful mechanism and even allowing non-separable CTRs
and we do so by introducing a new payment rule and utilizing results from \cite{AGM06} via the above reduction, and
(ii) where we restrict ourselves to the same ranking and payment rules but allow CTRs to be non-separable utilizing
results from \cite{BCPP06} via the above reduction. \\

It is known that the {\bf GSP} is not truthful\cite{AGM06,EOS05,Lah06} and clearly this holds true for {\bf Exp-GSP} as well.
And as we mentioned in the Section \ref{intro},
there is a result \cite{GP07} with a goal towards implementing
a truthful mechanism while learning the CTRs, and to achieve this goal it had to deviate from the current pricing scheme.
Our exploration based mechanism described in Section \ref{Exp-GSP} can also be made truthful by changing the payment rule.
All the description of the mechanism remains the same except the following:

\begin{itemize}
 \item The bidders are ranked by
$\tilde{q}_ib_i$ where $\tilde{q}_i$ is the quality score the search engines defines for the bidders $i$. For example, usual choices of
$\tilde{q}_i$ are search engines' estimate of $c_{i,1}$ or that of $\sum_{j=1}^K c_{i,j}$.

\item  The bidder $i$ is charged an amount per-click $p_i$ given by,
\begin{align}
p_i = \sum_{j=i}^{\tilde{K}} \frac{(\tilde{c}_{i,j}-\tilde{c}_{i,j+1})}{\tilde{c}_{i,i}} \frac{\tilde{q}_{j+1}b_{j+1}}{\tilde{q}_i}.
\end{align}
 \end{itemize}

In spirit of \cite{AGM06}, we call this variation of our exploration mechanism as {\bf Exp-Laddered} and it can be proved to be truthful
by adopting the proof in  \cite{AGM06}. We refer the usual one step truthful mechanism without any exploration
to as  {\bf Laddered}.  Now let us compute the {\it cost of uncertainty} in this truthful implementation
and as will see below we can obtain a similar upper bound as in Section \ref{revsec}.
Let $R_0$ be the expected revenue of the auctioneer for {\bf Laddered} and
$R$ be her expected revenue for {\bf Exp-Laddered} then
\begin{align}
R_0 = \sum_{i=1}^K \sum_{j=i}^K (c_{i,j}-c_{i,j+1}) \frac{\tilde{q}_{j+1}b_{j+1}}{\tilde{q}_i} \\
\nonumber \\
R = \sum_{i=1}^{\tilde{K}} \sum_{j=i}^{\tilde{K}} (\tilde{c}_{i,j}-\tilde{c}_{i,j+1}) \frac{\tilde{q}_{j+1}b_{j+1}}{\tilde{q}_i}
\end{align}

Performing calculations as in Section \ref{revsec}, we can obtain the following theorem.

\begin{theo}
\label{cout}
Let
\begin{equation}
c= \min_{1 \leq i \leq \min\{n,K\}} \min_{ i \leq j < n-L} \frac{c_{i,j+L} - c_{i,j+1+L}}{c_{i,j} -c_{i,j+1}}
\end{equation}
then the ``cost of uncertainty'' associated with truthful implementation is upper bounded by
\begin{align}
\left( 1 - \min\{1,c\} (1-\frac{2L}{n}) \right).
\end{align}
\end{theo}

Note that the Theorem \ref{cout} is consistent with Theorem \ref{cou} when we assume CTRs to be separable
i.e. $c_{i,j} =\gamma_j e_i$. \\

Now we consider the variation of {\bf Exp-GSP}
where we restrict ourselves to the same ranking and payment rules but allow CTRs to be non-separable.
If there were no restrictions on the ranking rule, following \cite{SS72,DGS86,BO06} we could argue that
there would always exist Walrasian equilibria and in particular such an equlibrium where every bidder
pays her opportunity cost. This equilibrium is called {\it MP pricing } equilibrium as
at this equilibrium every bidder obtains her marginal product as her payoff.
But there exists ranking rules for which there is no {\it MP pricing} equilibrium \cite{AGM06}.
As {\bf Laddered} is unique truthful mechanism given a weighted ranking rule, whenever
{\it MP pricing} equilibrium exists which is compatible with the ranking rule in {\bf Exp-GSP},
every bidder's payment is the same as in {\bf Exp-Laddered} and therefore
the expected revenue of the auctioneer at minimum {\em SNE} of {\bf GSP} and {\bf Exp-GSP} are same as
for {\bf Laddered} and {\bf Exp-Laddered} respectively. Thus the {\it cost of uncertainty} is
the same as in the case of truthful implementation and is given by Theorem \ref{cout}.
The existence of Walrasian equilibria (not necessarily the {\it MP pricing}) can be
explicitly proven for the ranking used in {\bf Exp-GSP} utilizing the
results from \cite{BCPP06},
but unfortunately it does not have a nice analytical form unlike in the seperable CTRs case or in the truthful case
and analytical computaton of {\it cost of uncertainty} does not seem feasible.
However, intuition from the earlier section indicates that similar results should hold as in Section \ref{revsec}. \\

It is clear that the estimation results from Section \ref{est} can easily be extended for both the variations
of {\bf Exp-GSP} discussed above and we omit the
detailed discussion.


\section{Concluding remarks}
\label{concl} We proposed a deterministic exploration mechanism to learn the relevance
scores by {\em deviating minimally} from the mechanism being currently used by Google and Yahoo!
in the sense that it retains the same pricing rule, as well as, similar ranking scheme. We
show that such an exploration policy can be incorporated without any significant loss in
revenue for the auctioneer. An independent work reported in \cite{WVLL07} introduces a
randomized exploratory mechanism and analyzes its incentive properties. We demonstrate
that the mathematical structure of the payoffs in our proposed  exploratory mechanism
({\bf EXP-GSP}) is identical to that in the standard mechanism (i.e., without
exploration), allowing us to compare and contrast the various metrics at the corresponding
\textbf{SNE}s. We show that while the actual bid profiles of {\bf Exp-GSP} and {\bf GSP}
may differ at the corresponding SNEs, the macroscopic measures, such as revenue,
efficiency etc. do not differ significantly, allowing auctioneers to limit the cost of
uncertainty. The approach in \cite{WVLL07}, on the other hand, centers around showing that
{\em both} the mechanisms (i.e., the standard GSP and the proposed exploratory randomized
mechanism ) would share \emph{almost-identical} equilibrium bid profiles; of course, the
auctioneer still pays a price for learning the quality factors (as in our case). These two
different approaches to the design of exploratory mechanisms raise an important topic for
future work: what other exploratory mechanisms can one design, and are their lower bounds
on the cost or price of uncertainty? That is, can one design mechanisms that have the
optimal characteristics when it comes to revenue loss vs. the information gathered about
quality factors and valuations.  Clearly, more work is necessary and more mechanisms such
as those proposed herein and in \cite{WVLL07} need to be studied.

\paragraph*{Acknowledgements:}
We thank Sushil Bikhchandani and Himawan Gunadhi for insightful discussions.
The work of SKS was partially supported by his internship at NetSeer Inc. Los Angeles.

\section*{Appendix}

{\bf Proof of Lemma \ref{thetas}:}
Let $m < L$, then
\begin{align*}
d_m= (n-L-(m-1))\gamma_{L+m} + \gamma_{L+1}+\gamma_{L+2}+\cdots+\gamma_{L+m-1} \\
d_{m+1}= (n-L-m)\gamma_{L+m+1} + \gamma_{L+1}+\gamma_{L+2}+\cdots+\gamma_{L+m} \\
\therefore d_m -d_{m+1} = (n - L -m) (\gamma_{m+L} - \gamma_{m+1+L})
\end{align*}
As we have $\gamma_j > \gamma_{j+1}$ for all $1 \leq j \leq K$, we get
\begin{align*}
d_m > d_{m+1}
\end{align*}
whenever $ m < n- L$ and $ m \leq K-L$ and therefore we have
\begin{align*}
d_1 > d_2 > \dots >d_{L-1} > d_L
\end{align*}
whenever $ L \leq \frac{1}{2} \min\{n,K+1\}$.

For $L \leq m < n-L$,
\begin{equation*}
\begin{array}{l}
d_m= (m-L)\gamma_m + \gamma_{m+1} + \dots + \gamma_{m+L-1}+(n-m-L+1) \gamma_{m+L}  \\
d_{m+1}= (m+1-L)\gamma_{m+1} + \gamma_{m+2} + \dots + \gamma_{m+L}+(n-m-L) \gamma_{m+L+1} \\
\therefore d_m -d_{m+1} =
(m-L) (\gamma_m - \gamma_{m+1})+(n-m-L)(\gamma_{m+L} - \gamma_{m+1+L})  \\
\\
d_L - d_{L+1} = (n-2L) (\gamma_{2L} - \gamma_{2L+1}) \\
 > 0 \textrm{  whenever  } n > 2L \textrm{  and  }  2L \leq K .
\end{array}
\end{equation*}
For, $L < m < n-L$,
clearly $(n-m-L)(\gamma_{m+L} - \gamma_{m+1+L}) \geq 0 $,
and $(m-L) (\gamma_m - \gamma_{m+1}) > 0$ whenever $m \leq K$ and therefore
$d_m > d_{m+1}$   whenever    $n \leq K+L+1$.
\begin{equation*}
\begin{array}{l}
\therefore d_L > d_{L+1} > \dots > d_{n-L} \\
\textrm{   whenever  }   L \leq \frac{1}{2} \min\{n-1,K\} \textrm{  and  }   n \leq K+L+1.
\end{array}
\end{equation*}
Further, for $n-L \leq m \leq n-1$,
\begin{align*}
d_m= (m-L) \gamma_m + \gamma_{m+1}+\dots+\gamma_n \\
d_{m+1}= (m+1-L) \gamma_{m+1} + \gamma_{m+2}+ \dots+\gamma_n \\
d_m -d_{m+1} = (m-L) (\gamma_{m} - \gamma_{m+1}) \\
\therefore d_m > d_{m+1} \textrm{   whenever  } m \leq K \\
\\
\therefore  d_{n-L} > d_{n-L+1} > \dots > d_n \textrm{   whenever  } n \leq K+1.
\end{align*}

Combining the above relations and noting that $\theta_j = \gamma + d_j$ for all $1 \leq j \leq n$, we obtain
\begin{align*}
\theta_j > \theta_{j+1}  \textrm{   for all  } 1 \leq j \leq n-1 \\
\textrm{   whenever  } L \leq \frac{1}{2} \min\{n-1,K\} \textrm{  and  }   n \leq K+1.
\end{align*}
Now,
$\theta_n -\theta_{n+1} = \gamma + (n-L) \gamma_n - n \gamma_{n+1} > 0$ whenever $ L > 0$ or $n \leq K$ and
for $j > n$ , $\theta_j -\theta_{j+1} = n (\gamma_j - \gamma_{j+1}) > 0$ whenever  $j \leq K$ and
$\theta_j -\theta_{j+1}$ is $0$ otherwise. This completes the proof.\qed

{\bf Proof of Theorem \ref{cou}:}

Now from proof of Lemma \ref{epctr}, we can observe that

$\theta_j - \theta_{j+1} =$
\begin{displaymath}
\left\{ \begin{array}{ll}
(n-j-L)(\gamma_{j+L} - \gamma_{j+1+L}) & ; j < L \\
(j-L)(\gamma_j - \gamma_{j+1})+(n-j-L)(\gamma_{j+L}-\gamma_{j+1+L}) & ; L \leq j < n-L \\
(j-L)(\gamma_j - \gamma_{j+1}) & ; n-L \leq j < n \\
(\gamma - L \gamma_{n+1}) + (n-L)(\gamma_{n} - \gamma_{n+1}) & ; j =n \\
n(\gamma_j - \gamma_{j+1}) & ; j>n\\
\end{array} \right.
\end{displaymath}

$\therefore \frac{\theta_j - \theta_{j+1}}{n(\gamma_j -\gamma_{j+1})}  = $
\begin{displaymath}
 \left\{ \begin{array}{ll}
(1-\frac{j+L}{n}) (\frac{\gamma_{j+L} - \gamma_{j+1+L}}{\gamma_j -\gamma_{j+1}}) & ; j < L \\
\frac{j-L}{n} +(1-\frac{j+L}{n}) (\frac{\gamma_{j+L} - \gamma_{j+1+L}}{\gamma_j -\gamma_{j+1}}) & ; L \leq j < n-L \\
\frac{1}{n} (j-L) & ; n-L \leq j < n \\
 \frac{1}{n} \frac{(\gamma - L\gamma_{n+1})}{(\gamma_n - \gamma_{n+1})} + (1-\frac{L}{n}) & ; j =n \\
1 & ; n < j \leq K
\end{array} \right.
\end{displaymath}
Let
\begin{align*}
c= \min_{1 \leq j < n-L} \frac{\gamma_{j+L} - \gamma_{j+1+L}}{\gamma_j -\gamma_{j+1}}
\end{align*}
\begin{align*}
\textrm{  then  }  \frac{\theta_j - \theta_{j+1}}{n(\gamma_j -\gamma_{j+1})}  \geq
\end{align*}
\begin{displaymath}
 \left\{ \begin{array}{ll}
(1-\frac{j+L}{n}) c & ; j < L \\
\frac{j-L}{n} +(1-\frac{j+L}{n}) c & ; L \leq j < n-L \\
\frac{1}{n} (j-L) & ; n-L \leq j < n \\
1-\frac{L}{n} & ; j =n \\
1 & ; n < j \leq K \\
\end{array} \right.
\end{displaymath}
\begin{displaymath}
\geq \left\{ \begin{array}{ll}
(1-\frac{2L}{n}) c & ; j < L \\
\\
(1-\frac{2L}{n}) \min\{1,c\} & ; L \leq j < n-L \\
\\
(1-\frac{2L}{n}) & ; n-L \leq j < n \\
\\
1-\frac{L}{n} & ; j =n \\
\\
1 & ; n < j \leq K
\end{array} \right.
\end{displaymath}
\begin{displaymath}
\geq  \begin{array}{ll}
(1-\frac{2L}{n}) \min\{1,c\} & ; 1 \leq j \leq K
\end{array}
\end{displaymath}
Therefore, for all  $1 \leq j \leq K$, we have
\begin{displaymath}
\begin{array}{l}
 (\gamma_j -\gamma_{j+1}) - \frac{1}{n} (\theta_j - \theta_{j+1}) \\
\\
\leq \left( 1 - \min\{1,c\} (1-\frac{2L}{n}) \right) ( \gamma_j -\gamma_{j+1} ).
\end{array}
\end{displaymath}
\begin{displaymath}
\begin{array}{l}
\therefore R_0- \frac{1}{n} R = \\
\sum_{s=1}^{\tilde{K}} \sum_{j=s}^{\tilde{K}} \frac{e_s}{q_s}  \left[(\gamma_j - \gamma_{j+1}) - \frac{1}{n}
(\theta_j - \theta_{j+1})\right] q_{j+1} v_{j+1} \\
\\
\leq \sum_{s=1}^K \sum_{j=s}^K \frac{e_s}{q_s}  \left[(\gamma_j - \gamma_{j+1}) - \frac{1}{n}
(\theta_j - \theta_{j+1})\right] q_{j+1} v_{j+1} \\
\\
\leq \sum_{s=1}^{\min\{n,K\}} \sum_{j=s}^{\min\{n,K\}} \frac{e_s}{q_s}  \left( 1 - \min\{1,c\} (1-\frac{2L}{n}) \right) ( \gamma_j -\gamma_{j+1} ) q_{j+1} v_{j+1} \\
\\
\leq \left( 1 - \min\{1,c\} (1-\frac{2L}{n}) \right) R_0^{\min\{n,K\}} , \\
\\
\textrm{ where  $R_0^l$ denotes the revenue of auctioneer from top $l$ bidders in {\bf GSP}} \\
\\
\therefore \frac{R_0- \frac{1}{n} R}{R_0} \leq \left( 1 - \min\{1,c\} (1-\frac{2L}{n}) \right) \left(\frac{R_0^{\min\{n,K\}}}{R_0}\right) \\
\\
 \leq \left( 1 - \min\{1,c\} (1-\frac{2L}{n}) \right) .
\end{array}
\end{displaymath}

{\bf Proof of Theorem \ref{effloss}:}

Using Lemma \ref{effre} we have,
\begin{displaymath}
\begin{array}{l}
E_0 - \frac{1}{n} E = \sum_{m=1}^K  \gamma_m e_m v_m - \frac{1}{n}  \sum_{m=1}^K \gamma_m y_m   \\
\\
 = \sum_{m=1}^K \gamma_m e_m v_m \left( 1 - \frac{1}{n} \frac{y_m}{e_m v_m}\right) .
\end{array}
\end{displaymath}

Let us first assume that
\begin{equation}
\label{evord}
e_m v_m \geq e_{m+1} v_{m+1} \textrm{  for all  }  1 \leq m \leq n.
\end{equation}

For $m \leq L$, we have
\begin{displaymath}
\begin{array}{l}
\frac{1}{n} \frac{y_m}{e_mv_m} = \frac{1}{n} \frac{\sum_{i=1}^n e_i v_i}{e_mv_m} \\
\\
\geq  \frac{1}{n}\frac{\sum_{i=1}^n e_i v_i}{e_1v_1} \\
\\
\therefore 1 - \frac{1}{n} \frac{y_m}{e_mv_m} \leq (1-\alpha) \\
\\
\textrm{   where   } \alpha = \frac{1}{n}\frac{\sum_{i=1}^n e_i v_i}{e_1v_1}.
\end{array}
\end{displaymath}
For $L < m \leq n$,
\begin{displaymath}
\begin{array}{l}
\frac{1}{n} \frac{y_m}{e_mv_m}= \frac{1}{n} \left[(n-m+1) (\frac{e_{m-L} v_{m-L}}{e_m v_m }) + \sum_{i=m-L+1}^{m-1} (\frac{e_iv_i}{e_m v_m} )
 + (m-L) \right]\\
\\
\geq \frac{1}{n} \left[(n-m+L) \left(\frac{e_{m-1} v_{m-1}}{e_m v_m }\right) + (m-L) \right]\\
\\
\therefore 1- \frac{1}{n} \frac{y_m}{e_mv_m} \leq \frac{1}{n} \left[(n-m+L) \left(1 - \frac{e_{m-1} v_{m-1}}{e_m v_m }\right)  \right]\\
\\
= - \frac{1}{n} \left[(n-m+L) \left( \frac{e_{m-1} v_{m-1}}{e_m v_m } - 1 \right)  \right]\\
\\
\leq - \frac{L}{n} \left(\frac{e_{m-1} v_{m-1}}{e_m v_m } - 1 \right) \\
\leq - \frac{L}{n} \omega \\
\end{array}
\end{displaymath}
where
\begin{align*}
\omega = \min_{L < m \leq n} \left( \frac{e_{m-1}v_{m-1}}{e_m v_m} - 1 \right).
\end{align*}
For $n < m \leq K $,
\begin{displaymath}
\frac{1}{n} \frac{y_m}{e_mv_m} = 1.
\end{displaymath}
Therefore,
\begin{align*}
E_0 - \frac{1}{n} E = \sum_{m=1}^L \gamma_m e_m v_m \left( 1 - \frac{1}{n} \frac{y_m}{e_m v_m}\right) \\
 +\sum_{m=L+1}^n \gamma_m e_m v_m \left( 1 - \frac{1}{n} \frac{y_m}{e_m v_m}\right) \\
+ \sum_{m=n+1}^K \gamma_m e_m v_m \left( 1 - \frac{1}{n} \frac{y_m}{e_m v_m}\right) \\
\\
\leq  (1-\alpha) \sum_{m=1}^L \gamma_m e_m v_m - \frac{L}{n} \omega  \sum_{m=L+1}^n \gamma_m e_m v_m \\
\\
= (1-\alpha) E_0^e - \frac{L}{n} \omega E_0^{ne} \\
\end{align*}
where  $E_0^{e} = \sum_{i=1}^L \gamma_m e_m v_m$ and $E_0^{ne} = \sum_{i=L+1}^{n} \gamma_m e_m v_m$.

But it might be the case that the Equation \ref{evord} does not hold.
In this case, we have
for $L < m \leq n$,
\begin{align*}
1 - \frac{1}{n} \frac{y_m}{e_mv_m} \leq (1-\beta) \\
 \textrm{ where    }
 \beta = \frac{1}{n}\frac{\sum_{i=1}^n e_i v_i}{\max_{1 \leq m \leq L} e_mv_m}
\end{align*}
and for $L < m \leq n$,
\begin{align*}
1- \frac{1}{n} \frac{y_m}{e_mv_m} \leq \eta \\
\\
\textrm{  where   }  \eta = \max_{L < m \leq n} \left\{\max_{m-L \leq i \leq m} \left(1-\frac{e_iv_i}{e_mv_m}\right)\right\}.
\end{align*}

\begin{align*}
\therefore E_0 - \frac{1}{n} E \leq (1-\beta) E_0^e + \eta E_0^{ne} \qed.
\end{align*}


\begin{thebibliography}{100}

\bibitem{AGM06} G. Aggarwal, A. Goel, R. Motwani, Truthful Auctions for Pricing Search Keywords, EC 2006.
\bibitem{BCPP06} T. Borgers, I. Cox,  M. Pesendorfer, V. Petricek, 
Equilibrium Bids in Sponsored Search Auctions: Theory and Evidence, Technical report, University of Michigan (2007).
\bibitem{BO06} S. Bikhchandani and J. M. Ostroy, From the assignment model to combinatorial auctions. In Combinatorial Auctions MIT Press 2006.
\bibitem{DGS86} Gabrielle Demange, David Gale, and Marilda Sotomayor, Multi-Item Auctions, Jour. Political Economy, 94, 863-872, 1986.
\bibitem{EOS05} B. Edelman, M. Ostrovsky,  M. Schwarz,
Internet Advertising and the Generalized Second Price Auction: Selling Billions of Dollars Worth of Keywords, American Economic Review 2007.
\bibitem{GP07} R. Gonen and E. Palkov, An Incentive-Compatible Multi-Armed Bandit Mechanism. In . Third Workshop on Sponsored Search Auctions WWW2007.
\bibitem{IJMT05} N. Immorlica, K. Jain, M. Mahdian, and K. Talwar,
Click Fraud Resistant Methods for Learning Click-Through Rates, WINE 2005.
\bibitem{Lah06} S. Lahaie, An Analysis of Alternative Slot Auction Designs for Sponsored Search, EC 2006.
\bibitem{LP07} S.  Lahaie, D. Pennock, Revenue Analysis of a Family of Ranking Rules for Keyword Auctions, EC 2007.
\bibitem{MNS07} M. Mahdian, H. Nazerzadeh, A. Saberi, Allocating online advertisement space with unreliable estimates, EC 2007
\bibitem{MSVV05} A. Mehta, A.   Saberi, U.  Vazirani, V.   Vazirani, AdWords and generalized on-line matching, FOCS 2005.
\bibitem{PO06} S. Pandey, and C. Olston, Handling advertisements of unknown quality in search advertising, NIPS 2006.
\bibitem{RDR07} M. Richardson, E. Dominowska, and R. Ragno, Predicting Clicks: Estimating the Click-Through Rate for New Ads, WWW 2007.
\bibitem{SBRR07} S. K. Singh, M. Bradonji\'c, V. P. Roychowdhury, and B. A. Rezaei,
Adword Auctions: Fairness Without Loss (available at http://arxiv.org/abs/0707.1053).
\bibitem{SS72} L. S. Shapley and M. Shubik, The Assignment Game I: The Core, Int. J. Game Theory 1, no. 2, 111-30, 1972.
\bibitem{TK91} A. Tversky; D. Kahneman, Loss Aversion in Riskless Choice: A Reference-Dependent Model,
The Quarterly Journal of Economics, Vol. 106, No. 4. (Nov., 1991), pp. 1039-1061.
\bibitem{Var06} H. Varian, Position Auctions, To appear in International Journal of Industrial Organization.
\bibitem{WVLL07} J. Wortman, Y. Vorobeychik, L. Li, and J. Langford,
Maintaining equilibria during exploration in sponsored search auctions, to appear in WINE 2007.


\end{thebibliography}
\end{document}